# Structural evolution of $CO_2$-filled pure silica LTA zeolite under high-pressure high-temperature conditions

David Santamaría-Pérez[1,*], Tomas Marqueño[1], Simon MacLeod[2,3], Javier Ruiz-Fuertes[1], Dominik Daisenberger[4], Raquel Chuliá-Jordan[1], Daniel Errandonea[1], Jose Luis Jordá[5], Fernando Rey[5], Chris McGuire[6], Adam Mahkluf[6], Abby Kavner[6], Catalin Popescu[7]

[1] MALTA-Departamento de Física Aplicada-ICMUV, Universidad de Valencia, 46100, Valencia, Spain.
[2] Atomic Weapons Establishment, Aldermaston, Reading, RG7 4PR, UK
[3] Institute of Shock Physics, Imperial College London, London, SW7 2AZ, UK
[4] Diamond Light Source, Didcot OX11 0DE, Oxon, UK
[5] Instituto de Tecnología Química, Universitat Politècnica de València – Consejo Superior de Investigaciones Científicas, 46022, Valencia, Spain.
[6] Earth, Planetary and Space Sciences Department, University of California Los Angeles, 951567, Los Angeles, USA.
[7] ALBA-CELLS, 08290, Cerdanyola, Spain

**Abstract**

The crystal structure of $CO_2$-filled pure-$SiO_2$ LTA zeolite has been studied at high pressures and temperatures using synchrotron-based x-ray powder diffraction. Its structure consists of 13 $CO_2$ guest molecules, 12 of them accommodated in the large α-cages and 1 in the β-cages, giving a $SiO_2$:$CO_2$ stoichiometric ratio smaller than 2. The structure remains stable under pressure up to 20 GPa with a slight pressure-dependent rhombohedral distortion, indicating that pressure-induced amorphization is prevented by the insertion of guest species in this open framework. The ambient-temperature lattice compressibility has been determined. In situ high-pressure resistive-heating experiments up to 750 K allow us to estimate the thermal expansivity at P~5 GPa. Our data confirm that the insertion of $CO_2$ reverses the negative thermal expansion of the empty zeolite structure. No evidence of any chemical reaction was observed. The possibility of synthesizing a silicon carbonate at high temperatures and higher pressures is discussed in terms of the evolution of C – O and Si – O distances between molecular and framework atoms.

**Introduction**

The long-standing quest of novel compounds within the $CO_2$:$SiO_2$ system has recently taken a step forward after considering simultaneous high-pressure (HP) and high-temperature (HT) conditions [1-5]. In particular, a recent study reported the synthesis of a silicon carbonate phase by reacting a well-known $SiO_2$ zeolite, silicalite, with the molecular $CO_2$ that fills the pores at 18-26 GPa and 600-980 K [1]. The large effective surface exposed to the $CO_2$ in the pores within this [$SiO_4$]-based compound was likely a crucial factor for enhancing the chemical reaction. Raman and infrared spectroscopy measurements of temperature-quenched samples reveal the formation of unidentate and bidentate carbonate units involving one or two framework silicon atoms [1], in a similar way of $CO_2$ adsorbed at ambient pressure on the surface of basic metal oxides [6,7]. In situ HP-HT x-ray diffraction measurements only show a progressive broadening of the silicalite diffraction peaks upon increasing temperature [1]. These results suggest the existence of novel $CO_2$-$SiO_2$ chemistry systematics with a wealth of uncharacterized compounds, and with potentially interesting chemical properties. Several theoretical total-energy studies have predicted novel silicon carbonate phases at HP-HT and their results need to be experimentally confirmed [8-10]. Note that the supposedly advantage of using zeolites to maximize surface chemical reactivity due to the large effective interaction area between the framework $SiO_2$ and the confined-$CO_2$ seems to be restricted to temperatures below 1300 K [3,4]. Above that temperature, zeolites transform into the thermodynamically stable phase of $SiO_2$ at the corresponding pressure, from quartz to stishovite. At present, the amount of $CO_2$ accommodated in the pores of open framework structures, the exact location of these $CO_2$ molecules and the structural and chemical behavior of these systems under high pressures and high temperatures are still open questions that remain to be answered.

There exist a large number of synthetic porous silica structures [11]. Zeolites are a class subdivision consisting of structures with channel-like voids with windows large enough for guest species to pass through [12,13]. The selection of the zeolite structural type for seeking $CO_2$-$SiO_2$ reactivity should be based on three criteria: (i) high symmetry and structural simplicity (few $SiO_2$ formula units per unit cell), (ii) the proved absence of aluminum or any cations in the structure, and (iii) large cavities or cages to host a considerable number of $CO_2$ molecules. Pure silica zeolite A, named ITQ-29, is an ideal candidate. It has an LTA structure, with a cubic Pm-3m space group and only 24 formula units per unit cell, which can be described as a three-dimensional network of spherical, 1.14 nm cavities (α-cages), interconnected by six small windows that are limited by eight-membered ring pores with a diameter of 0.41 nm (see Fig. 1)

[14]. An alternative description of this structure is in terms of sodalite cages (β-cages) which are connected through square faces with each other, forming double four-ring units [14]. The adsorption of $CO_2$ by this relatively simple zeolite structure has been recently studied by means of molecular dynamic simulations [15], the results showing that the density of $CO_2$ is uniform within the α-cages, but denser in the 8-ring window between the cages. These data suggest a longer residence time in these preferred locations, with the carbon atom of the carbon dioxide molecule in the center of the window and the oxygens pointing out into the cages on either side.

The present study aims to locate and subsequently evaluate the potential chemical interaction of adsorbed $CO_2$ molecules in the aforementioned ITQ-29 host framework. Here, we report in situ angle-dispersive synchrotron x-ray diffraction (XRD) data on the $CO_2$-filled ITQ-29 system at high-pressure and high-temperature conditions. Phase transitions have been identified, relevant compressibility-expansivity parameters determined and its structural behavior discussed.

**Experimental details**

The ITQ-29 sample, which is the pure-silica form of zeolite A, was prepared as previously described in literature [12], and calcined at 1173 K for 8 h to remove all organic substances. Its structure was confirmed by in-home XRD patterns and had a density of 1.435 g/cm$^3$.

We performed three kind of experiments: (i) at room pressure (RP) and HT, (ii) at HP and room temperature (RT), and (iii) at simultaneous HP-HT conditions. Several facilities and pressure cells were used in the different experiments performed. The RP-HT XRD measurements were performed using an Anton-Paar XRK-900 reaction chamber attached to a PANalytical X'Pert PRO diffractometer. For RT compression experiments to 20 GPa, symmetric diamond-anvil cells (DACs) were equipped with diamonds of 350 µm. The ITQ-29 sample was placed together with a small amount of Pt metal at the center of a 100 µm diameter hole in the rhenium gasket, preindented to a thickness of 40 µm. The equation of state (EOS) of Pt was used as a pressure gauge [16], which provided pressure accuracies of 0.1 GPa below 10 GPa and 0.2 GPa up to 20 GPa. Subsequently, high-purity $CO_2$ was loaded in the DAC at room temperature using the COMPRES gas loading apparatus. In situ HP-RT angle-dispersive X-ray diffraction measurements were carried out at synchrotron-based beamline GSECARS at the Advanced Photon Source. At GSECARS, incident X-rays had a wavelength of 0.4592 Å, the beam was focused to ~5x5 µm$^2$ spot, and diffraction patterns measured from 5 to 40 seconds were collected on a CCD camera.

For the HP resistive-heating studies we used three gas-membrane-driven DACs equipped with diamonds with 300 μm culets and loaded with the same sample in the same manner as described above. NaCl powder was used as the pressure marker [17]. $CO_2$ was loaded in the DAC at room temperature using the Sanchez Technologies gas loading apparatus available at Diamond Light Source. The DACs were contained within a custom-built vacuum vessel designed for HP–HT experiments [18,19]. The DACs were heated using Watlow 240V (rated at 4.65W cm$^{-2}$) coiled heaters wrapped around the outside of the DACs. The temperature was measured using a K-type thermocouple attached to one of the diamond anvils, close to the gasket. The accuracy of the thermocouple on the temperature range covered by the experiments is 0.4%. Pressure uncertainties are of the same order as those in room-temperature experiments. The diffraction data were collected using the MSPD beamline at ALBA-CELLS synchrotron using a x-ray wavelength of 0.534 Å (Rh K edge) [20]. The beamline is equipped with Kirkpatrick-Baez mirrors, to focus the X-ray beam to ~20x20 μm$^2$, and with a Rayonix CCD detector.

A precise calibration of the detectors parameters was developed with reference $LaB_6$ and $CeO_2$ powder, and integration to conventional $2\theta$-intensity data were carried out with the Dioptas software [21]. The indexing and refinement of the powder patterns were performed using the Fullprof [22] and Powdercell [23] program packages.

**Results**

**A. Room-pressure high-temperature experimental data**

The evolution of the lattice parameter of the empty $SiO_2$-pure ITQ-29 zeolite at room pressure and high temperature has been studied by means of XRD measurements. Data at 10 different temperatures in the range 293 – 1173 K are collected in Table 1. The linear thermal expansion coefficient $\alpha_V/3$ is found to be negative, varying from -5.55 × 10$^{-6}$ to -4.55 × 10$^{-6}$ K$^{-1}$ over the aforementioned temperature range. This fact has been previously observed in many pure silica zeolites and the driving force of the contraction mechanism has been attributed to changes in the ∠Si-O-Si intertetrahedral bond angles [24,25]. In agreement with previous studies [25], we suggest that the negative thermal expansion is likely due to transverse vibrations of the bridging oxygen atoms of the open structure.

## B. High-pressure room-temperature experimental data

Figure 2 shows a selection of the diffraction patterns at different pressures and Figure 3 shows the caked images of three raw CCD images, to illustrate the quality of the XRD data. At 0.5 GPa, the structure can be described as a cubic Pm-3m LTA-type phase previously reported in the literature [14] with comparable lattice parameter: a = 11.8293(5) Å, as obtained from a LeBail refinement. Atomic coordinates were initially not refined since the exact $CO_2$ content and the position of the hosted $CO_2$ molecules is unknown. However, the existence of carbon dioxide in the cages can be inferred from (i) the change in peak intensity ratios between empty and the observed zeolite patterns, and (ii) as it will be seen later, the large stability of the host framework structure upon compression. The lattice parameters of the zeolite could have increased between 0.6 and 1.1% with $CO_2$ adsorption, as inferred from the RP-RT unit cell volume data of the empty zeolite (see Table 1) and the zero-pressure volume estimated from the RT filled-$CO_2$ zeolite EOS (see next paragraph). In fact, if we assume that the atomic coordinates of the host have not changed significantly between the empty structure at ambient pressure and the supposedly $CO_2$-filled structure at 0.5 GPa, the whole structure at this latter pressure could be completed by Rietveld refinement and Fourier recycling [26]. Thus, the electron density map, derived from the contributions of the [$SiO_4$] units of the zeolite only in a difference Fourier map ($F_{obs} - F_{calc}$), should allow to tentatively locate some $CO_2$ molecules in the porous structure. Nevertheless, in spite of the high quality powder data and the simplicity of the concept, the determination of the exact location of $CO_2$ molecules was difficult in practice. The only information that could be extracted was the existence of extra peaks in planes z=0, that would correspond to the (0, 0, 0) coordinates of the sodalite β-cage center and the (0.5, 0.5, 0) coordinates of the α-cage windows, in planes z=0.31 and 0.69, that would correspond to the (0.31, 0.31, 0.31) coordinates, and in planes z=0.5, that would correspond to the (0.5, 0.5, 0.5) coordinates of the α-cage center (see Figure 4). We used those positions as initial locations of the center of gravity of the $CO_2$ molecules, the C atoms. Subsequently, the O atoms were placed at known 1.17 Å distances using the Fourier maps and the zeolite topology as reference (see Fig. 1b). The atomic coordinates so-determined are collected in Table 2. The insertion of these thirteen $CO_2$ molecules considerably improves the observed-calculated peak intensity ratios and the overall model refinement (see Fig. 5). Note that this adsorption value is larger than that uptaken by several Al-containing LTA zeolites (6-9 $CO_2$ molecules/unit cell) at 0.5 MPa and 303 K, as calculated from $CO_2$ adsorption isotherms [27]. The location of these molecules defines the interatomic distance between the C atom of carbon dioxide and the nearest O atoms of the host silica framework, 2.98 Å (compare to the intramolecular C-O distance, 1.17 Å), which is one of the relevant parameters to be tuned at high pressures and temperatures to enhance chemical reactivity and carbonate formation. The other one, the closest distance between Si atoms and O atoms of the $CO_2$ molecule is approx. 3 Å.

This structure is stable to 1.4 GPa at which point the low-angle (110) and (111) diffraction peaks progressively broaden as a consequence of a reflection splitting due to symmetry reduction. Above 5 GPa, the peak splitting is clearly visible, confirming the existence of a structural phase transition (see Figure 2). The diffraction peaks (12 lines) could be indexed in a rhombohedral unitcell with lattice parameters at 5.4 GPa: a = 11.48(2) Å, α = 91.0(4)°. The space group R-3m (Nr. 166) could explain the observed reflections and correspond to a translationengleiche subgroup of the initial Pm-3m group. Therefore, the phase transition taking place is a second-order displacive transition. This is supported by the fact that no discontinuity in the unit cell volume was observed at the transition. The evolution of the volume per formula unit and the rhombohedral angle of the unit cell with increasing pressure are illustrated in Figure 6. Data are collected in Table 3. The pressure-volume data in the quasi-hydrostatic range (P≤ 12 GPa, 5 P-V points) were analyzed using a third-order Birch-Murnaghan equation of state (EOS), with three fitting parameters: The zero-pressure volume ($V_0$), the bulk modulus ($B_0$) and its first pressure derivative ($B'_0$). For $CO_2$-filled ITQ-29 we find $V_0$ = 1679(9) Å$^3$, $B_0$ = 39(3) GPa, and $B'_0$ = 3.3(3). As the amount of data are very limited, we also fit a second-order Birch-Murnaghan EOS, with characteristic parameters $V_0$ =1687(2) Å$^3$ and $B_0$ = 38.0(3) GPa. A similar compressibility behavior has been reported for $CO_2$-filled silicalite, with a bulk modulus of $B_0$ = 35.5 GPa and a $B'_0$ = 4.0 [28,29]. Note that the observed compressibility in these filled zeolites is comparable to that of α-quartz [30], $B_0$ = 37.12(9) GPa and $B'_0$ = 5.99(4), which indicates direct compression of the framework.

The volume variation with pressure described by the aforementioned EOS explains all our XRD data from both APS and ALBA experiments except for a single data point measured at ALBA at 6.3 GPa. Its volume is larger than that expected for this pressure, suggesting a smaller compressibility (see Figure 6). This could be caused by either a deterioration of the quasi-hydrostatic conditions produced by loading a smaller amount of $CO_2$ in the pressure chamber in the ALBA experiment, or the use of a different pressure standard calibrant. As shown in the next section, the estimation of this slightly different compressibility behavior will be crucial for obtaining thermal expansivity data. Thus, we fitted the ALBA XRD data to a 2$^{nd}$-order Birch-Murnaghan EOS and obtained a bulk modulus $B_0$ = 51(2) GPa.

The Pm-3m to R-3m transformation is basically caused by the compression along the diagonal of the cubic unit cell, which produces its distortion. No structural refinements could be performed at high pressures due to the limited quality of the diffracted signal and the uncertainty in the position of the $CO_2$ guest molecules. However, assuming that the atomic coordinates of the $SiO_2$ framework remain fixed, the process would entail small displacements of the Si and O atoms from their high-symmetry cubic positions in the host structure, producing a slight tilting of the rigid [$SiO_4$] tetrahedra around bridging oxygen atoms [31,32], and likely the relocation of the $CO_2$ molecules within the cages. This movement would explain the change of intensities in the diffraction patterns above the transition pressure (see Figure 2). This phase seems to be stable up to 19.9 GPa, the maximum pressure reached in the present study, although the bad quality of the patterns at the highest pressures precludes an unequivocal indexation. Nevertheless, the observation of low 2θ-angle x-ray diffraction lines at high pressures evidences the absence of pressure-induce amorphization (PIA), which is consistent with complete filling of the pores by $CO_2$ in an equilibrium process with the rest of $CO_2$ outside the pores that acts as pressure transmitting medium (PTM).

Two recent studies have performed high-pressure XRD and neutron diffraction experiments on pure-silica ITQ-29 using a non-penetrating PTM, like silicone oil and arsenolite [33,34]. Under compression, ITQ-29 undergoes two transformations at 1.2 and 3.2 GPa, respectively, the first one fully reversible and the second one, non-reversible upon pressure release [33]. This latter phase was fully characterized as a novel zeolite type, ITQ-50, with different cage topology to ITQ-29. Neither of these phases was observed in our experiments, likely due to the occupation of channels by penetrating $CO_2$ that limits the flexibility of the structure and prevents collapse of the framework.

### C. High-pressure high-temperature experimental data

The high-pressure high-temperature chemistry of the $CO_2$-$SiO_2$ system was recently studied using double-sided laser heating in diamond-anvil cells, while characterizing the samples *in situ* by means of synchrotron-based X-ray diffraction. Samples included the $CO_2$-filled ITQ-29 zeolite together with $CO_2$ pressure transmitting medium and a metallic heater, which acted as absorber of the laser. Upon heating the compressed sample (up to 50 GPa) above 1300 K, it is shown that the zeolite transformed into the thermodynamically stable phase of $SiO_2$ at the corresponding pressures, quartz for $P < 4$ GPa and stishovite for $P > 8$ GPa [3,4]. Therefore, the supposedly advantage of maximizing chemical reactivity due to the large effective interaction area between the framework $SiO_2$ and the confined-$CO_2$ is restricted to lower temperatures. Thus, in order to evaluate the possible chemical interaction between host and guest species we used externally resistive-heating DAC experiments (up to 830 K) and characterized *in situ* by means of synchrotron x-ray radiation. In our experimental run, the sample was initially compressed to 6.3 GPa, far beyond the cubic-to-rhombohedral transition of the zeolite, subsequently heated to 635 K, further compressed and heated to 9.6 GPa and 825 K, and finally isothermally compressed up to 24.3 GPa and temperature quenched to ambient temperature. The P-T path followed in our experiment is shown in Figure 7, together with the phase diagrams of $CO_2$ [35] and $SiO_2$ [36].

The pressure in the sample chamber decreases slightly with increasing temperature, from 6.3 GPa at ambient temperature to 4.4 GPa at 587 K (see HP-HT data collected in Table 4). In this P-T range, the $CO_2$ used as PTM remains solid, in phase I, and the distorted rhombohedral $CO_2$-filled ITQ-29 structure is stable. Above that temperature, the $CO_2$ PTM becomes liquid as shown in its phase diagram in Figure 7 and all the $CO_2$ diffraction maxima disappeared as a consequence of the loss of long-range order in melting. At that point, the pressure in the sample chamber drops 1.2 GPa with only 60 K temperature increase and the structure of the $CO_2$-filled ITQ-29 framework recover the initial cubic symmetry. Thus, this zeolite structure seems to be extremely sensitive to surrounding non-hydrostatic conditions and even a soft solid media like the $CO_2$-I molecular phase produces a structural distortion along the diagonal of the unitcell. In fact, as depicted in Figure 8, the splitting of the reflections slightly decrease with increasing temperature (the clearer example is the (311) cubic reflection) and, only close to 630 K, the rhombohedral angle rapidly moves to 90° and the structure becomes cubic. This HP-HT phase transition data point together with that of the Pm-3m to R-3m transition at HP-RT conditions allows us to tentatively constrain the phase boundary as a straight line with positive Clapeyron slope of 7.8 MPa/K (see Figure 7).

Because our measured volume data of $CO_2$-filled ITQ-29 zeolite in both, the cubic and the rhombohedral structures, are clustered around relatively narrow pressure and temperature ranges, and because the pressure is not constant during heating of the sample (Figure 7), we used a perturbational method [37] to extract thermal expansivity at ~5 GPa, using the room-temperature equation of state for ITQ-29 as a reference (see previous Section). Starting with a first order Taylor expansion of volume in terms of pressure and temperature, and the definitions of isothermal bulk modulus ($K_T$) and thermal expansivity (α), we can estimate the latter by subtracting the reference volume at each pressure, and normalizing to the measured volume:

$$\frac{V_{P,T} - V_{ref}}{V_{P,T}} = -\frac{(P - P_{ref})}{K_T} + \alpha(T - T_{ref})$$

In our analysis, we use the extrapolated Birch-Murnaghan equation of state for ITQ-29 as a reference curve. Therefore, the slope of a plot of $1 - \dfrac{V_{ref}}{V_{meas}}$ versus ($T - T_{ref}$) is a measure of the volumetric thermal expansivity (Figure 9). A linear fit to our high P-T volumes yields a thermal expansion value of $6.6(2) \cdot 10^{-5}$ K$^{-1}$. This is a robust conclusion from our data, since the linearity of the plot suggests that the first order Taylor expansion is an appropriate approximation. The positive linear expansivity of the $CO_2$-filled silica structure at ~5 GPa contrasts to the negative linear expansivity of the empty zeolite at ambient pressure, which suggests that guest molecules minimize [SiO$_4$] tetrahedral tilting with increasing temperature.

Upon further compression (4.8 GPa) and heating (737 K), the $CO_2$-filled ITQ-29 silica zeolite transforms into the thermodynamically stable phase of $SiO_2$ at these conditions, coesite. The diffraction peaks of the XRD patterns are perfectly explained by the lattice parameters and coordinates found in literature (see Table 5) [38,39]. This phase transition allows us to tentatively draw a ITQ-29–to–coesite phase boundary line, which sets an upper bound on the HP-HT stability of this silica pure zeolite. At higher pressures, coesite transforms into stishovite as expected.

**Discussion**

Assuming that reactivity would mainly occur at zeolite cavities, the possible reaction pathways between $CO_2$ accommodated in cages and the host $SiO_2$ structure would depend on their stoichiometric ratio, which ultimately determines the final location of the guest $CO_2$ molecules and the interatomic distances. Our results suggest that 13 molecules of $CO_2$ are present in the large α cavities (12) and the sodalite β cavities (1) of the LTA structure (Z=24), providing a lower content bound of $CO_2$:$SiO_2$ > 0.5. The α-cages are connected with neighboring cages through the 4.1 Å diameter windows formed by the [SiO$_4$] eight-membered rings, which ensures that $CO_2$ molecules can easily move. Moreover, under pressure conditions of 0.5 GPa, $CO_2$ is also able to penetrate inside the sodalite cages through the six-membered ring windows. Our observed stochiometric ratio is much smaller than the 1:1 considered recently in an *ab initio* molecular dynamics simulations study on the related RHO-type silica-pure zeolite [40]. This is expected since RHO structure is formed solely by α-cages that can accommodate $CO_2$, while the LTA structure contains α-cages and sodalite cages [41]. $CO_2$ access to the latest cavities is more limited, resulting in a lower adsorption capacity. In that study the authors simulated the effect of high pressures and high-temperatures in the chemical interaction between $SiO_2$ and $CO_2$. At 800 K, the simulated compression on the system put the different atoms closer and predicts a major structural change at approx. 44 GPa, that involves a sudden change of the atom's coordination. The C atoms that were mostly in the sp-hybridation below that pressure are bonded to three or four O atoms above it, whereas tetrahedrally-coordinated Si atoms also increase their number of O neighbor atoms up to five or six. In other words, $CO_2$ and $SiO_2$ react forming an amorphous silicon carbonate. According to simulations using the 1:1 stochiometric ratio, transition occurs when the $C_{CO2}$ – $O_{SiO2}$ and Si – $O_{CO2}$ distances are of the order of 2.3 and 2.55 Å, respectively [40], which is not far from 2.61 and 2.66 Å we can estimate at 19.9 GPa assuming that atomic coordinates did not change upon compression. These interatomic distances decrease rapidly with increasing pressure by means of the lattice parameters contraction in a highly compressible porous framework. It seems clear that pressures P>30 GPa, well above those considered in our HP-HT study, are required to bring closer and enhance the chemical activity of the $CO_2$ molecules and the host silica framework. At those pressures, the additional C – O and Si – O interactions will likely give rise to electron redistribution that could eventually lead to an increase in coordination and the formation of a silicon carbonate.

Besides compression, high temperatures could also favor chemical reactivity due to stronger motions of atoms and molecules and seems to be crucial in this system [1]. This was suggested by molecular simulations [40], which predicted that the fraction of ∠O-C-O angles at 120º would increase significantly at 1300 K (and 32 GPa). However, this possibility was recently ruled out by high-pressure laser-heating experiments on two zeolite types, silicalite and ITQ-29, which confirmed that porous structures undergo phase transformations into the thermodynamically stable phases of silica below 1300 K. The results of the present high-pressure resistive-heating study seem to be in excellent agreement with previous experimental observations, the $CO_2$-filled ITQ-29 structure transforming at ~4 GPa into coesite + $CO_2$ below ~730 K. Further experiments at higher pressures (5<P<60 GPa) and moderate temperatures (T<1300 K) using spectroscopic techniques are needed to explore the potential synthesis of silicon carbonate phases from porous silica and carbon dioxide.

**Conclusions**

We have studied the structural behavior of $CO_2$-filled ITQ-29 LTA zeolite under high-pressure and high temperature conditions, and our results provide an overview of the phase diagram of this

$CO_2$:$SiO_2$ system. The usual compression mechanism in LTA zeolite observed in high-pressure experiments using non-penetrating pressure media [33,34], which involves the collapse of the structure around the empty pores and the formation of the denser ITQ-50 zeolite at 3.2 GPa, is hindered by the presence of guest $CO_2$ molecules. Our results indicate that 13 molecules of carbon dioxide per formula unit were accommodated in the α- and β-cages of this silica-pure zeolite structure and that their insertion leads to the deactivation of the PIA mechanism, the ITQ-29 material being retained in a slightly distorted rhombohedral form to pressures of at least 20 GPa. X-ray powder diffraction data at low pressures allow us to determine the location of the guest molecules in the cage, which is a critical step for understanding the existing host-guest atomic interactions. Thermodynamic conditions were subsequently tuned to try to enhance the chemical activity of the different species. Room-temperature compression data up to 20 GPa yield the isothermal equation of state and high-temperature data allow us to determine an estimation of the thermal expansivity at P~ 5 GPa. These results suggest that reactivity between $CO_2$ and $SiO_2$ in this system would only occur above pressures of 30 GPa and temperatures below 1300 K, where zeolite would transform into stishovite, the thermodynamically stable phase of silica at that pressures. Spectroscopic experiments (ITQ-29 progressively losses the long-range order with increasing pressure) are needed to evaluate the possibility of synthesizing silicon carbonates from porous silica frameworks and embedded $CO_2$ molecules.

## Acknowledgements

The authors thank the financial support of the Spanish Ministerio de Economıa y Competitividad (MINECO), the Spanish Research Agency (AEI) and the European Fund for Regional Development (FEDER) under Grants No. MAT2016-75586-C4-1-P, MAT2015-71842-P, Severo Ochoa SEV-2012-0267 and No.MAT2015-71070-REDC (MALTA Consolider). D. S-P. and J. R-F acknowledge MINECO for a Ramón y Cajal and a Juan de la Cierva contract, respectively. Portions of this work were performed at GeoSoilEnviroCARS (Sector 13), Advanced Photon Source (APS), Argonne National Laboratory. GeoSoilEnviroCARS is supported by the National Science Foundation - Earth Sciences (EAR-1128799) and Department of Energy- GeoSciences (DE-FG02-94ER14466). This research used resources of the Advanced Photon Source, a U.S. Department of Energy (DOE) Office of Science User Facility operated for the DOE Office of Science by Argonne National Laboratory under Contract No. DE-AC02-06CH11357. Use of the COMPRES-GSECARS gas loading system was supported by COMPRES under NSF Cooperative Agreement EAR 11-57758. $CO_2$ gas was also loaded at Diamond Light Source.

Authors thank synchrotron ALBA-CELLS for beamtime allocation at MSPD line. British Crown Owned Copyright 2017/AWE. Published with permission of the Controller of Her Britannic Majesty's Stationery Office.

Table 1.- Lattice parameters and unit cell volumes for $SiO_2$ pure ITQ-29 zeolite at ambient pressure and different temperatures.

| Temperature (K) | Unitcell parameter (Å) | Unitcell volume (Å$^3$) |
|---|---|---|
| 303 | 11.8590(3) | 1667.78(8) |
| 373 | 11.8544(3) | 1665.84(8) |
| 473 | 11.8487(3) | 1663.45(8) |
| 573 | 11.8419(3) | 1660.59(8) |
| 673 | 11.8367(4) | 1658.40(9) |
| 773 | 11.8318(4) | 1656.43(9) |
| 873 | 11.8264(4) | 1654.10(9) |
| 973 | 11.8215(4) | 1652.03(9) |
| 1073 | 11.8171(4) | 1650.18(9) |
| 1173 | 11.8124(4) | 1648.21(9) |

Table 2.- Atomic coordinates of the $CO_2$-filled ITQ-29 structure at 0.5 GPa. The Si and O positions of the host structure (4 first lines) were fixed to those experimentally observed at ambient conditions. The atomic positions of the $CO_2$ molecules were inferred from maxima in Fourier maps, fixing the intramolecular C – O distance to 1.17 Å. Space group: Pm-3m. Lattice parameter: a = 11.8293(5) Å.

| Atom | Wyckoff position | x | y | z | Occupancy Factor |
|---|---|---|---|---|---|
| Si | 24k | 0 | 0.6299 | 0.1833 | 1 |
| O1 | 12h | 0.2217 | 0.5 | 0 | 1 |
| O2 | 12i | 0 | 0.2969 | 0.2969 | 1 |
| O3 | 24m | 0.1115 | 0.1115 | 0.3435 | 1 |
| C1 | 8g | 0.689 | 0.689 | 0.689 | 1 |
| O4 | 24m | 0.619 | 0.619 | 0.689 | 0.333 |
| O5 | 24m | 0.759 | 0.759 | 0.689 | 0.333 |
| C2 | 3c | 0 | 0.5 | 0.5 | 1 |
| O6 | 6f | 0.099 | 0.5 | 0.5 | 1 |
| C3 | 1b | 0.5 | 0.5 | 0.5 | 1 |
| O7 | 6f | 0.599 | 0.5 | 0.5 | 0.333 |
| C4 | 1a | 0 | 0 | 0 | 1 |
| O8 | 6e | 0.901 | 0 | 0 | 0.333 |

Table 3.- Lattice parameters of the $CO_2$-filled ITQ-29 zeolite at high-pressures and ambient temperature. Data measured at APS and ALBA-CELLS data are differentiated.

| Pressure (GPa) | a axis (Å) | α angle (º) | Unitcell Volume (Å$^3$) |
|---|---|---|---|
| 0.5 (APS) | 11.8293(5) | 90.0 | 1655.6(2) |
| 5.4 (APS) | 11.482(2) | 91.0(1) | 1512.3(5) |
| 7.5 (APS) | 11.351(2) | 91.4(1) | 1460.8(5) |
| 8.8 (APS) | 11.271(2) | 91.6(1) | 1429.7(5) |
| 12 (APS) | 11.108(3) | 91.9(1) | 1369.0(8) |
| 14.6 (APS) | 10.961(5) | 92.1(1) | 1313.8(14) |
| 17.6 (APS) | 10.822(9) | 92.4(2) | 1263(3) |
| 19.9 (APS) | 10.72(2) | 92.5(2) | 1228(7) |
| 0.4 (ALBA-CELLS) | 11.8280(11) | 90.0 | 1654.8(3) |
| 0.5 (ALBA-CELLS) | 11.8268(11) | 90.0 | 1654.3(3) |
| 0.8 (ALBA-CELLS) | 11.8079(10) | 90.0 | 1646.4(3) |
| 2.6 (ALBA-CELLS) | 11.667(2) | 90.4(1) | 1587.9(6) |
| 6.1 (ALBA-CELLS) | 11.478(4) | 91.2(2) | 1512.2(14) |

Table 4.- Lattice parameters of the $CO_2$-filled ITQ-29 zeolite at high-pressures and high-temperatures.

| Pressure (GPa) | Temperature (K) | a axis (Å) | α angle (º) | Unitcell Volume (Å$^3$) |
|---|---|---|---|---|
| 6.35 | 298 | 11.492(5) | 91.2(2) | 1517(2) |
| 6.20 | 337 | 11.518(5) | 91.1(2) | 1527.1(14) |
| 6.12 | 381 | 11.526(4) | 91.1(2) | 1530.3(15) |
| 6.10 | 423 | 11.535(4) | 91.1(1) | 1534.0(11) |
| 5.85 | 459 | 11.555(3) | 91.0(1) | 1542.1(8) |
| 5.70 | 502 | 11.579(4) | 91.0(1) | 1551.7(11) |
| 5.29 | 539 | 11.610(3) | 90.8(1) | 1564.6(8) |
| 4.42 | 587 | 11.675(2) | 90.5(1) | 1591.2(5) |
| 3.67 | 621 | 11.727(2) | 90.4(1) | 1612.6(5) |
| 3.30 | 646 | 11.737(1) | 90 | 1616.9(3) |
| 3.55 | 653 | 11.733(1) | 90 | 1615.2(3) |
| 3.85 | 699 | 11.723(1) | 90 | 1611.1(4) |

Table 5.- Lattice parameters of coesite at high-pressures and high-temperatures.

| Pressure (GPa) | Temperature (K) | a axis (Å) | b axis (Å) | c axis (Å) | β angle (º) | Unitcell Volume (Å$^3$) |
|---|---|---|---|---|---|---|
| 4.78 | 736 | 7.02(3) | 12.231(4) | 7.12(3) | 120.6(4) | 525(3) |
| 5.92 | 769 | 6.99(3) | 12.211(4) | 7.10(3) | 120.7(4) | 521(3) |
| 6.95 | 800 | 6.96(2) | 12.182(5) | 7.09(2) | 120.8(2) | 516(2) |
| 7.50 | 813 | 6.94(2) | 12.166(6) | 7.08(2) | 120.8(3) | 513(2) |
| 8.20 | 819 | 6.93(4) | 12.146(8) | 7.08(3) | 120.9(5) | 513(5) |
| 9.96 | 820 | 6.90(3) | 12.118(5) | 7.06(4) | 120.9(4) | 506(4) |

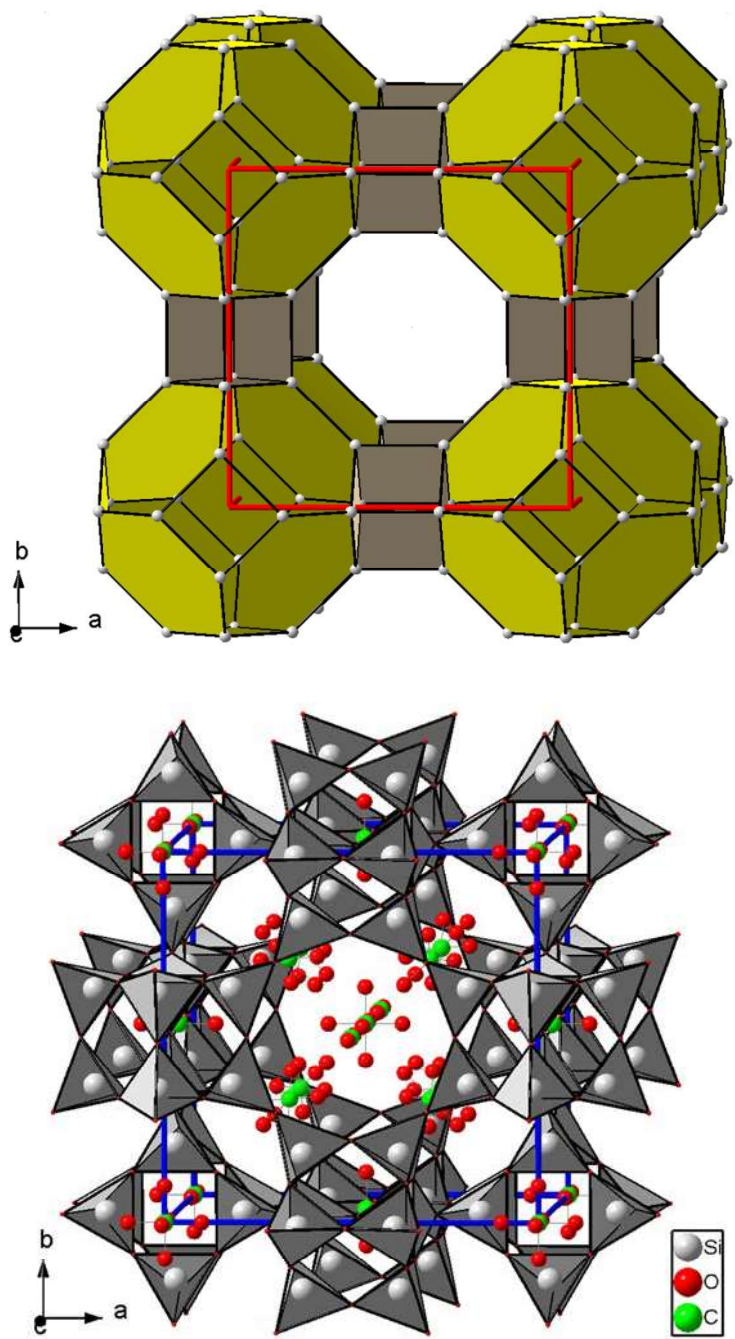

Figure 1.- (a) Packing of polyhedra formed by the host [SiO$_4$]-based framework in ITQ-29 zeolite. [4$^6$6$^8$] cubooctahedra and [4$^6$] cubes are depicted in yellow and grey, respectively. (b) The same framework structure showing the [SiO$_4$] connectivity and the location of CO$_2$ molecules at 0.5 GPa.. Grey, green and red spheres represent Si, C, and O atoms, respectively.

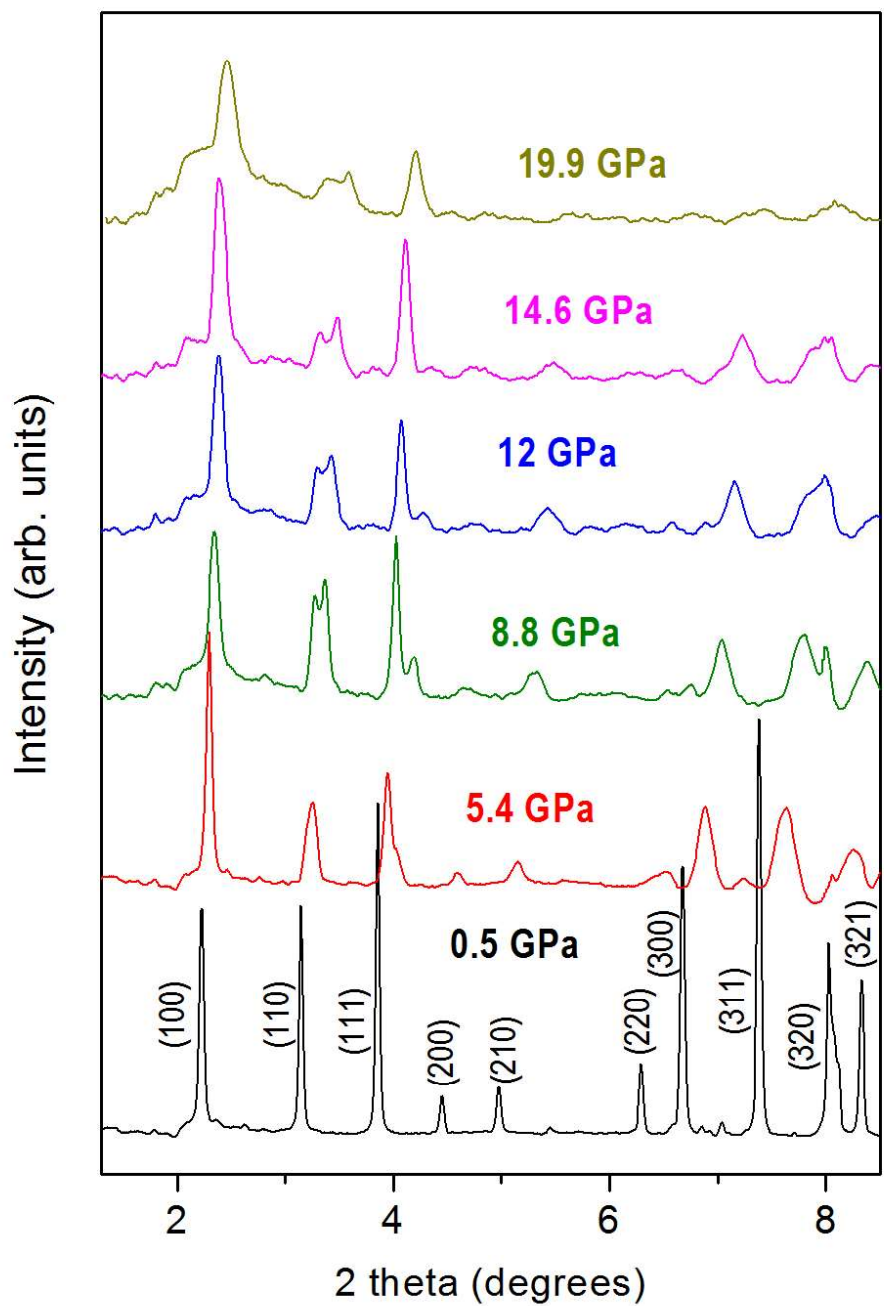

Figure 2.- Selected powder XRD patterns of $CO_2$-filled ITQ-29 pure silica zeolite up to 19.9 GPa. Backgrounds subtracted. Low-angle Bragg peaks of the porous structure are still clearly visible at the highest pressure.

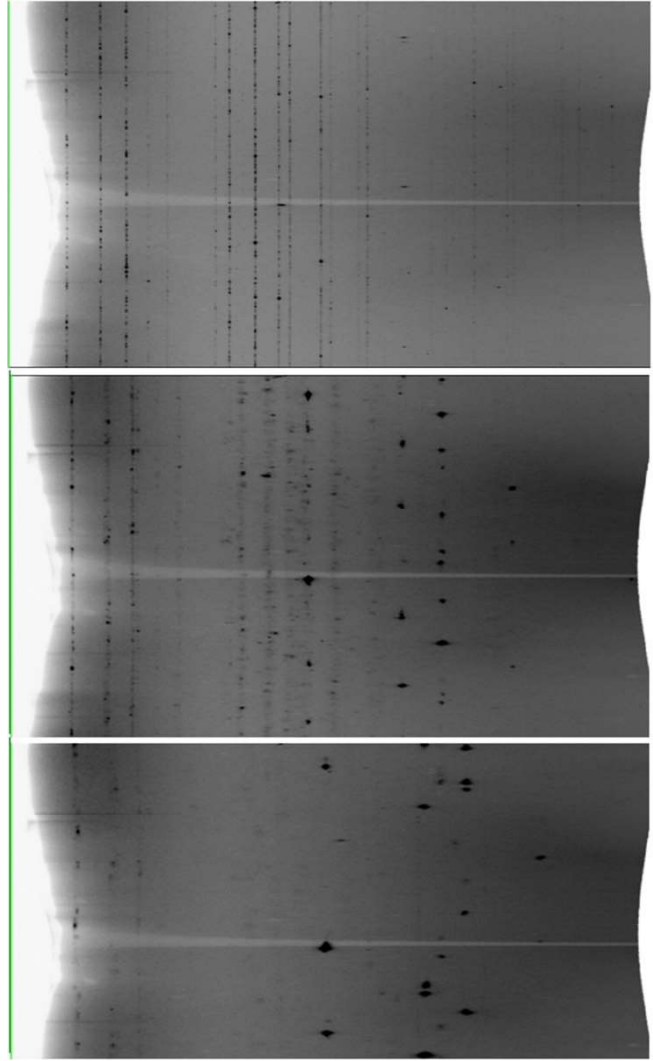

Figure 3.- Unrolled (or cake) images of the raw diffraction image from CCD detector at 0.5 GPa (top), ~5.4 GPa (center) and ~14.6 GPa (bottom) in the DAC. Images were processed using DIOPTAS program [15]. The cake images show differences in diffraction textures, information that is not recorded when using a standard integrated pattern.

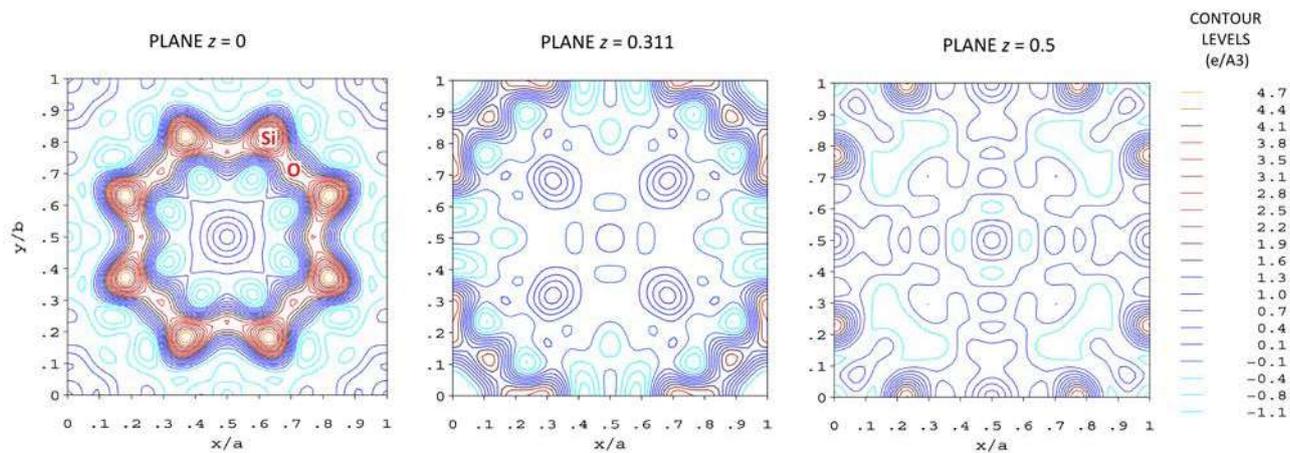

Figure 4.- Bidimensional Fourier map of the planes $z = 0$, $z = 0.311$ and $z = 0.5$ in $CO_2$-filled ITQ-29 zeolite calculated using x-ray structure factors.

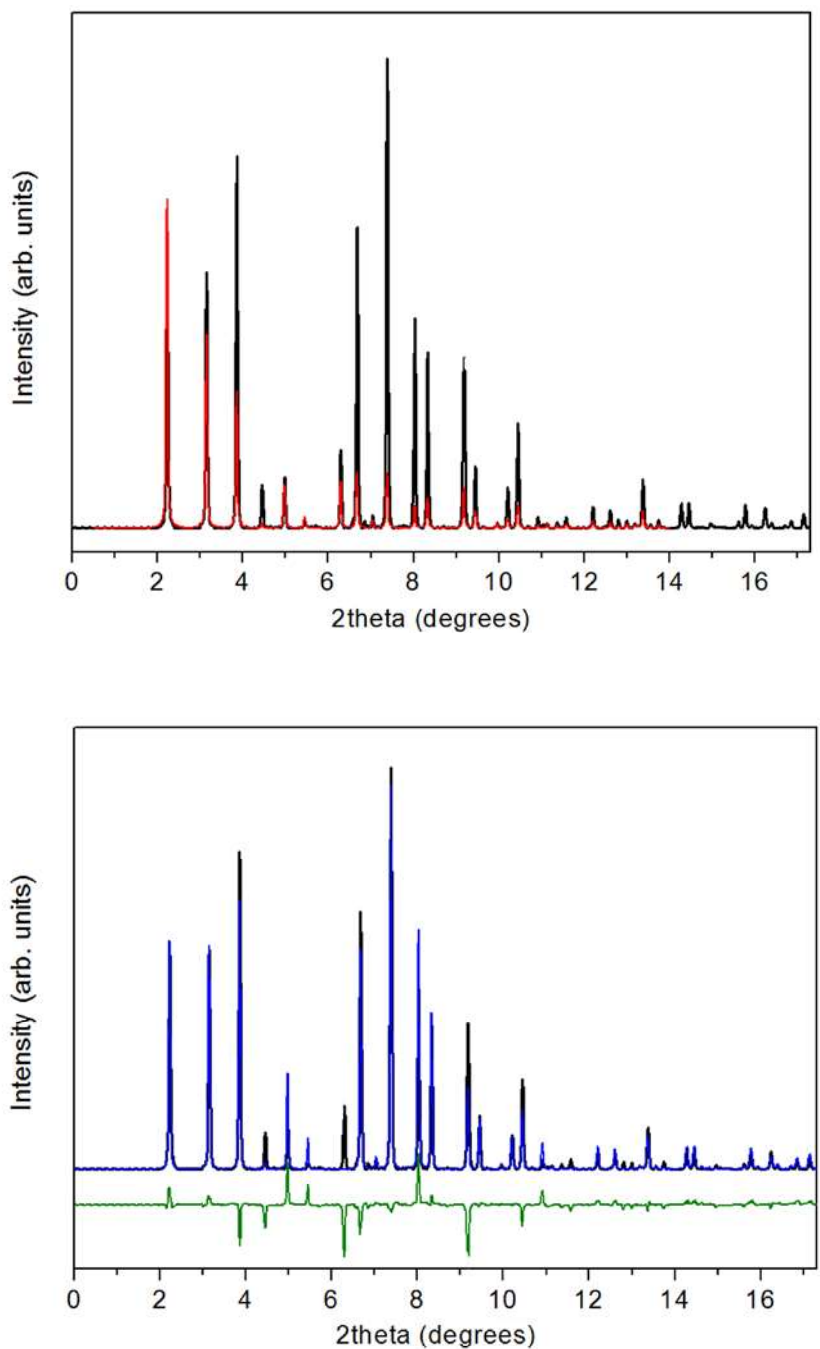

Figure 5.- XRD data of our sample ITQ29 at 0.5 GPa in black is compared to simulated models corresponding to the empty ITQ29 structure (above) and the ITQ29 structure filled with 13 $CO_2$ molecules (below). In the case of the empty structure, the misfit is not depicted for the sake of clarity (at 2θ > 6°, the misfit is of the same intensity order as the observed peaks).

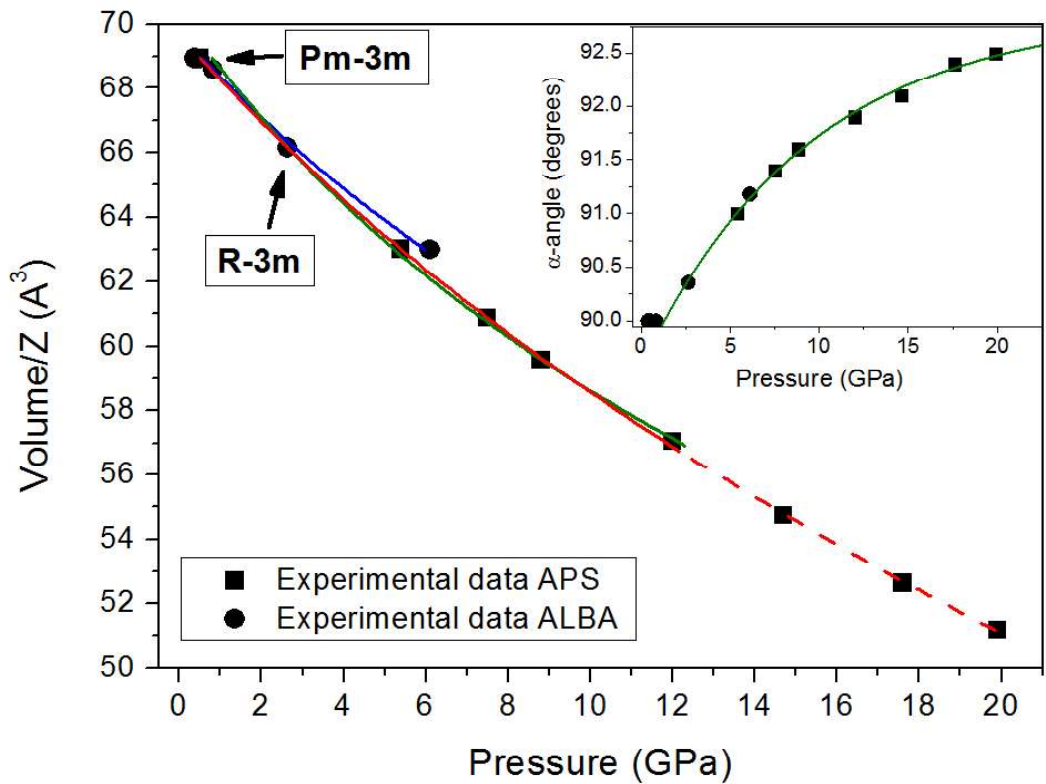

Figure 6.- P-V data per formula unit. Square and circle symbols refer to experimental data measured in APS and ALBA synchrotrons, respectively. Error bars are smaller than symbol size. The red solid line represents the fit of experimental data at P<12 GPa with a third-order Birch–Murnaghan EOS (excluding the 6.3 GPa data point measured in ALBA). The red dashed line shows the extrapolation of the fit to higher pressures. The green and blue solid lines are the second-order Birch–Murnaghan EOS fits to the APS and ALBA data, respectively. Inset: Evolution of the rhombohedral (α) angle with pressure.

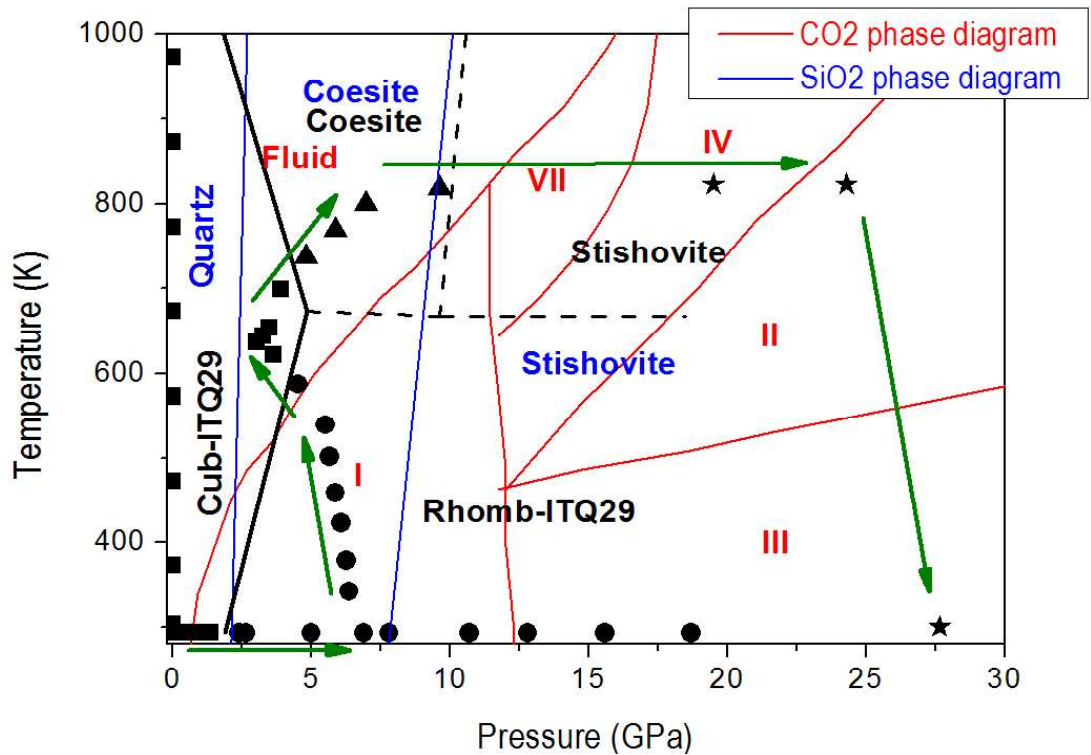

Figure 7.- P-T data for the ITQ-29 sample. Symbols in the vertical and horizontal axes correspond to RP-HT and HP-RT data. The path of the HP-HT run is denoted by green arrows. Solid squares, circles, triangles and stars correspond to the cubic Pm-3m ITQ-29, the rhombohedral R-3m ITQ-29, coesite and stishovite $SiO_2$ phases, respectively. Solid and dashed black lines denote tentative phase boundaries. Solid blue and red lines correspond to phase boundaries in $SiO_2$ and $CO_2$ phase diagrams.

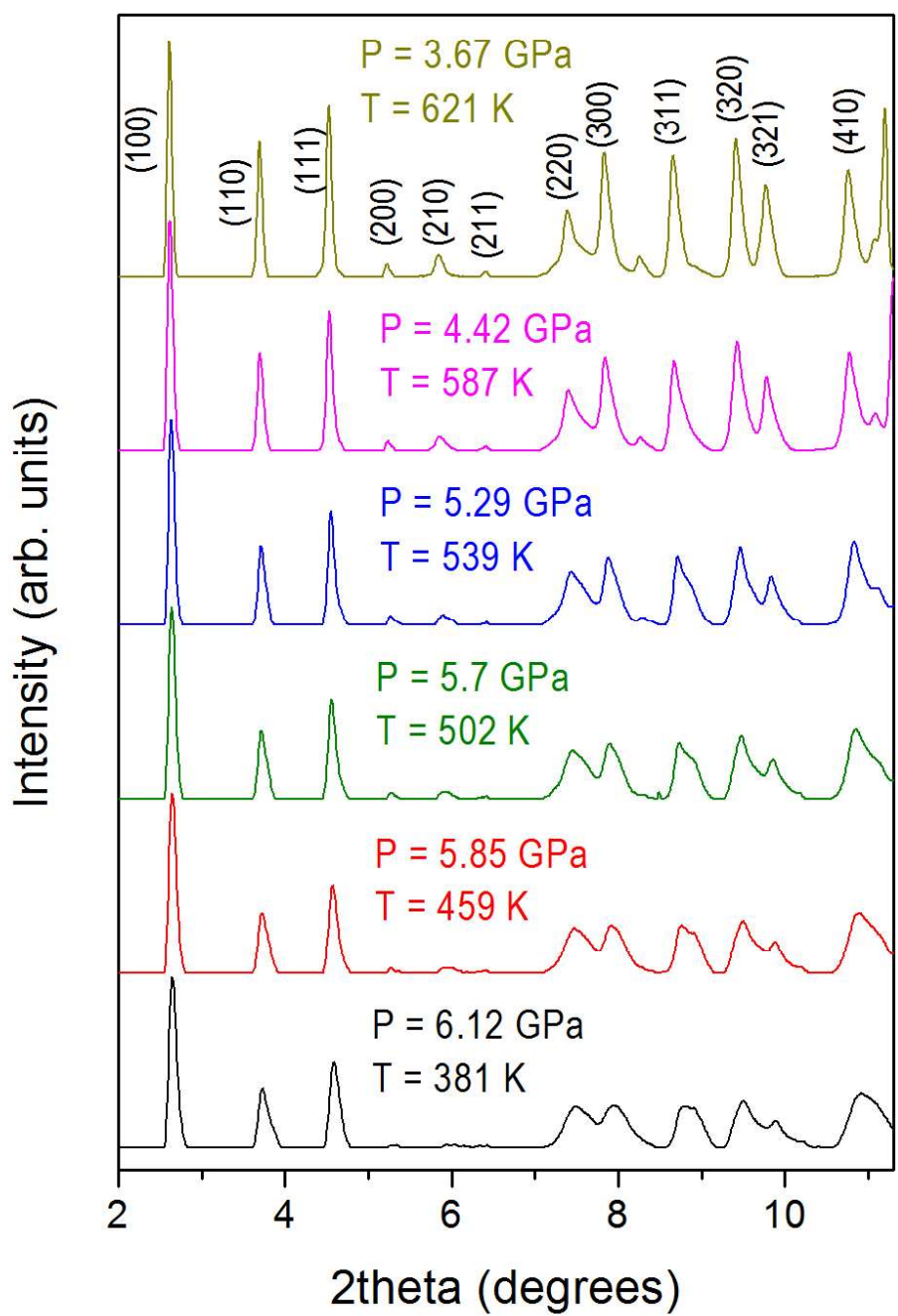

Figure 8.- Selected powder XRD patterns of the rhombohedrally-distorted $CO_2$-filled ITQ-29 zeolite at high-pressure (P~5 GPa) and high temperatures (up to 621 K). Backgrounds and kapton contribution subtracted. Peaks splitting clearly decrease with increasing temperature, indicating reduction in distortion until the cubic structure is formed (at ~ 621 K).

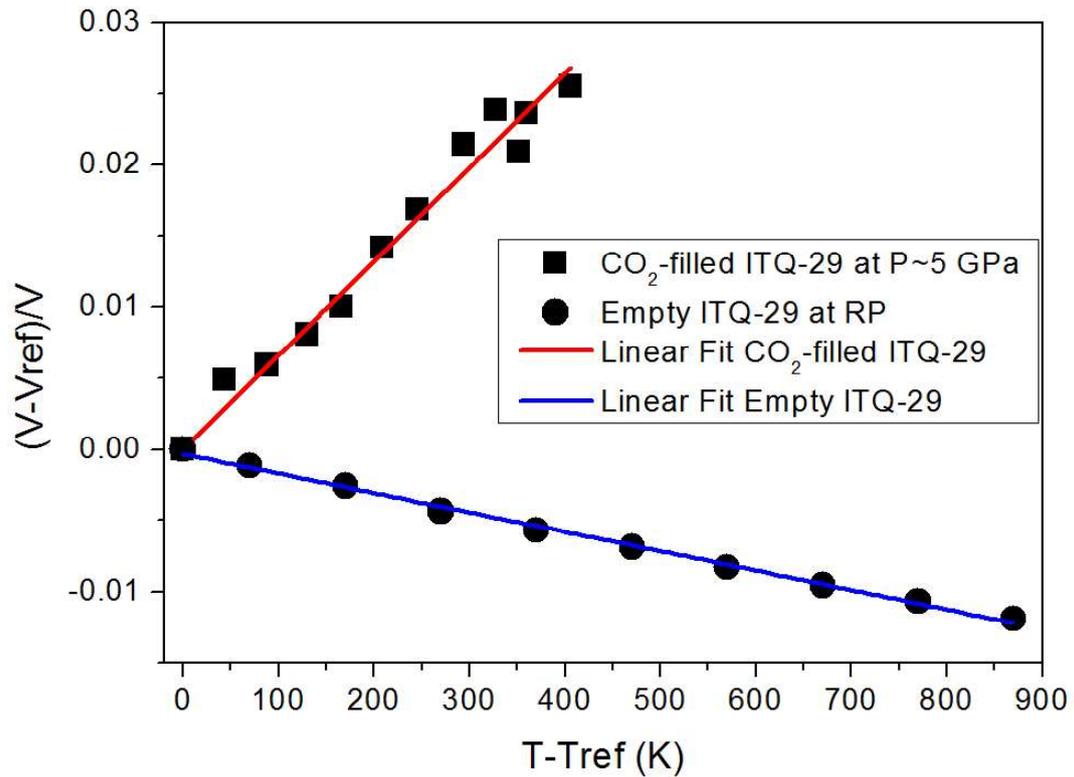

Figure 9.- Squares: Difference between measured high-temperature volume for $CO_2$-filled ITQ-29 and room-temperature reference volume (RT-EOS), divided by the measured volume, at each pressure as a function of temperature. The slope of the linear fit (red line) provides an estimation of the average thermal expansion for $CO_2$-filled zeolite at this average pressure. Circles and the blue line show the negative thermal expansivity of the empty structure at ambient pressure, for the sake of comparison.